\documentclass[%
 reprint,
  superscriptaddress,
showpacs,
showkeys,
preprintnumbers,
nofootinbib,
 amsmath,amssymb,
aps,
 prd,
 ]{revtex4}

\usepackage{graphicx}
\usepackage{dcolumn}
\usepackage{bm}
\usepackage{hyperref}

\usepackage{subfigure}
\usepackage{multirow}

\usepackage{epsf}
 \usepackage{color}

\def\beq{\begin{equation}}
\def\be{\begin{equation}}
\def\eeq{\end{equation}}
\def\ee{\end{equation}}
\def\bea{\begin{eqnarray}}
\def\eea{\end{eqnarray}}

\newcommand{\gsim}{\mbox{${~\rightarrowise.25em\hbox{$>$}\kern-.70em
\lower.25em\hbox{$\sim$}~}$}}
\newcommand{\lsim}{\mbox{${~\rightarrowise.25em\hbox{$<$}\kern-.70em
\lower.25em\hbox{$\sim$}~}$}}

\newcommand{\etp}{\eta ^{\prime}}

\def\ra{\rightarrow}
\def\D0{D\O }

\begin{document}


\title{Semileptonic $D$ decays and $\eta-\eta^\prime$ mixing
}


 \bigskip

 \author
{Giulia Ricciardi} 
\email{giulia.ricciardi@na.infn.it}
\affiliation
{Dipartimento di Scienze Fisiche, Universit\`a di Napoli Federico II \\
Complesso Universitario di Monte Sant'Angelo, Via Cintia, 80126 Napoli,
Italy}
\affiliation
{INFN, Sezione di Napoli, \\
Complesso Universitario di Monte Sant'Angelo,  Via Cintia, 80126 Napoli,
Italy}

\begin{abstract}

We discuss the impact of recent  progress in semileptonic $D$ decays
  on light flavor spectroscopy and estimate  mixing parameters in  the  $\eta-\eta^\prime$ system. 

\end{abstract}



   \maketitle
\flushbottom

In the past decade, charm decays have not received the same first-rate attention as beauty decays, but they are rapidly gaining ground.
A great shove  has been   the first evidence for $CP$  violation  in neutral $D$ meson decays, provided by LHCb and confirmed by CDF,
which has displayed  a size of the  asymmetry, almost a percent,  unexpectedly large. 
Detailed and comprehensive analyses of charm
transitions provide us with new
insights into  nonperturbative dynamics of QCD. Besides, they can help  calibrate
 theoretical tools for $B$ studies. Experimental facilities such as $BABAR$  and  CLEO have produced copious amounts of data, while  
 BESIII and LHCb are still collecting larger and larger samples of charmed hadrons.
The future flavor factories will probably  pursue charm measurements to
their ultimate precision.

We analyze the impact of  some of the latest progress in semileptonic charm decays on the $\eta-\eta^\prime$ system.
Determining the composition   of the $\eta$ and $\eta^{\prime}$ wave functions  is a long-standing problem, and a 
large number of phenomenological studies have been performed, based  on the  investigation of several different
 processes.
 At low energies,
radiative vector and pseudoscalar meson decays,
$\eta^{(\prime)}$  decays into two photons, or production in $\gamma\gamma$ collisions, two body 
decays of $\psi$ into $\eta^{(\prime)}$ plus a vector meson
have been carefully analyzed.
The results are not always in agreement,  leaving  unsolved issues, 
such as the possibility of gluonic mixing, and  motivate further investigation.

 Semileptonic $D_s^+ \rightarrow \eta^{(\prime )} l^+ \nu$ and  $D^+\rightarrow \eta^{(\prime )} l^+ \nu$ decays can   be a useful probe  for  $\eta-\eta^{\prime}$, in analogy with
 semileptonic or hadronic $B$ decays (see e.g. 
  \cite{MyBReport, DiDonato:2011kr, Fleischer:2011ib}).
Data on charm semileptonic decays   have existed since
1995, when  CLEO extracted the branching fractions  ${\cal{B}} ( D_s^+ \rightarrow \eta^{(\prime )} e^+ \nu_e)$ 
from ratios to hadronic decays of the $D^+_s$ \cite{cleo}. 
In 2009  the first absolute
measurement of  ${\cal{B}} ( D_s^+ \rightarrow \eta^{(\prime )} e^+ \nu_e)$ \cite{:2009cm} and 
 the first observation of the $ D^+ \rightarrow \eta \, e^+ \nu_e$ decay \cite{Mitchell:2008kb} were reported by the same collaboration.
Improved  branching
fraction measurements, together with the first observation of the decay mode $ D^+ \rightarrow \eta^\prime e^+ \nu_e  $  and
the first form factor determination for $ D^+ \rightarrow \eta \,  e^+ \nu_e$,   followed in 2011 \cite{Yelton:2010js}.
Recent experimental and theoretical progress has increased the role of semileptonic $D$ decays.

Here we exploit two major advances, the experimental measurements of all the relevant branching ratios and the lattice direct evaluation of the form factor shapes. 
We perform a phenomenological analysis of opportune ratios, aimed at estimating the $\eta-\eta^\prime$ mixing angle in a systematic way,
 keeping approximations under control. 



The simplest parametrization of the  mixing, including  the $gg$ component, can be expressed  
  in the heavy quark basis in terms of two mixing angles $\phi$ and $\phi_G$.  Estimates based on QCD sum rules 
 suggest that the coupling of the gluonium to the $\eta^\prime$ is larger than its coupling to $\eta$ \cite{DeFazio:2000my},
 which is also mainly an $SU(3)_{fl}$ octet. By
neglecting for simplicity the $g g$ component  in the $\eta$  state, we can write
\begin{eqnarray}
|\eta^{\prime}\rangle &\simeq & \cos  \phi_G \sin  \phi |\eta_q \rangle +\cos  \phi_G
\cos \phi|\eta_s \rangle + \sin \phi_G |gg \rangle \nonumber \\
|\eta\rangle &\simeq & \cos   \phi |\eta_q\rangle - \sin  \phi |\eta_s \rangle
\label{colla1}
\end{eqnarray}
in the quark-flavor basis, where the quark content of the isoscalar nonstrange and strange wave functions, assumed with
the same radial component, are $|\eta_q\rangle =\frac{1}{\sqrt 2}|u \bar u + d \bar d\rangle $ and
$|\eta_{s} \rangle = |s \bar s\rangle $.

In the past few years, several lattice results have  become available for the values of mixing angles \cite{Christ:2010dd, Dudek:2011tt}, 
 at about a decade of distance from the first estimate by UKQCD \cite{McNeile:2000hf}.
They are all in agreement,   quoting values of $\phi$ between
$  40^\circ$ and $50^\circ$, with errors slightly lower than the ones from phenomenological determinations.
The latest  analysis, by ETM,
leads to a value of
 $\phi  = (44 \pm 5)^\circ$ \cite{Ottnad:2012fv},
with statistical error only.  Systematic uncertainties, difficult to estimate on the lattice, are likely to affect this result.
Preliminary  results by the QCDSF Collaboration \cite{Bali:2011yx, Kanamori} give  a mixing angle $\theta \sim -(7^\circ, 8^\circ)$ in the octet-singlet basis, that is, in the quark-flavor basis,  $\phi = \theta+\arctan \sqrt{2} \sim 47^\circ $.
 Out of chorus is the lower value favored by  the recent UKQCD staggered investigation \cite{Gregory:2011sg}, 
 $\phi  = (34 \pm 3)^\circ$.
All lattice analyses do not include a gluonic operator, discussing only the
relative quark content.

Mixing angles extracted from  lattice QCD should  be compared with theoretical determinations from  phenomenology, which also allow estimates of the gluonic content.
Phenomenological  analysis of semileptonic decays  presented in  \cite{DiDonato:2011kr} results in  $\phi \sim 40^\circ$,  with theoretical and experimental uncertainties that are comparable and  total   errors ranging from 2\% to 10\%.
The  estimate of $\phi_G$  is at the moment dominated by the experimental errors that  prevent any conclusion on the gluonic content of the
$\eta - \eta^\prime$
system. There is clearly room for improvement on both the experimental and theoretical sides.

In this work, we estimate the values of  mixing  angles from  semileptonic $D$ decays, employing recently available data and some status of art theoretical determinations of form factors. The extraction from semileptonic $D$  decays
 has the potentiality to become increasingly more important, e.g. 
employing  new data we expect from  BESIII or from future Super Flavor factories.

In the past literature, the  knowledge of the mixing angle was sometimes used to compensate for  missing estimates, by  relating
 $ D_{(s)}^+ \rightarrow \eta e^+ \nu_e  $ and $ D_{(s)}^+ \rightarrow \eta^\prime e^+ \nu_e  $ decays. For instance, in QCD sum rule calculations of form factors \cite{Colangelo:2001cv} only the form factor 
$F_1^{ D_s \ra \eta}$ is directly available,  and $F_1^{D_s \ra \eta^\prime}$  is 
obtained, given $\phi$,  from the ratio
\beq
\frac{\left|F_1^{ D_s \ra \eta}(q^2)\right|}{\left|F_1^{D_s \ra \eta^\prime}(q^2)\right|} = \tan \phi
\label{rough}
\eeq
derived directly from Eqs. \ref{colla1} in the absence of gluon contributions.
However, as shown in \cite{DiDonato:2011kr}, estimates of the angle $\phi$ may suffer from a residual dependence on the model and the processes used for its extraction, not easy to quantify,  the simplest example being the variation due to the inclusion or the exclusion of the mixing with the gluonic component.
As of today, newly available  shapes of the form factors  allow one to extract directly the mixing angle values from semileptonic $D$ decays.
Preliminary  lattice determinations for $F_{1,0}(q^2)$   have been provided by the HPQCD \cite{Koponen:2012fu} and for  $F_0(q^2)$    by the  QCDSF Collaboration \cite{Bali:2011yx}.   Disconnected diagrams, which may account for mixing with gluonium,  have been taken into account in \cite{Bali:2011yx} only, and 
their effect  has been  found not negligible. More precise calculations are awaited, but we can already obtain  preliminary phenomenological  estimates of the mixing angles. A first immediate value  comes by comparing  at $q^2=0$ the  lattice determination of
 $F_1(0)=F_0(0) \sim [0.72-0.78] $ \cite{Koponen:2012fu}  with the QCD sum rule  calculation of $F_1^{ D_s \ra \eta}(0) =0.50 \pm 0.04$ \cite{Colangelo:2001cv}.
The form factor $F_1(q^2)$ calculated on lattice  \cite{Koponen:2012fu} does not refer to the physical meson, but to a pseudoscalar lattice 
construction made of a strange quark and antiquark, since only  the connected contributions to the $ D$ meson semileptonic decay form factor have been calculated.  It is found to be almost the same as  the form factor  $F_1(q^2)$  for the $D \ra K$ decay and at each $q^2$ it can approximately represent the form factor of transition to the $\eta_s$ part of the $\eta-\eta^\prime$ wave functions.
 The form factor  $F_1^{ D_s \ra \eta}(0)$ in  \cite{Colangelo:2001cv} refers to the physical state; therefore, according to Eqs. \ref{colla1} we find
\beq
 \frac{\left|F_1^{ D_s \ra \eta}(0)\right|}{\left|F_1^{D_s \ra \eta_s}(0)\right|} = \sin \phi
\label{rough1}
\eeq
which gives $\phi \sim 41^\circ$ with an error of about 8 \% . It is remarkable that,
 in the basic previous approximations and  by using two different theoretical approaches, we obtain the right order of magnitude of the mixing angle, with an error that is large, but comparable with errors from other determinations at lower energy, e.g. $\phi$  extractions employing  $\psi \ra \rho/\omega/\phi + \eta^{(\prime)}$ two body strong decays.

The  WA (Weak Annihilation) process may affect semileptonic meson decays \cite{Bigi:1993bh, Bianco:2003vb}. 
The observed differences in $D^{\pm,0}$ and $D_s$ semileptonic widths \cite{Asner:2009pu} may   in part be originated by  the valence spectator quark contributions in $D_s$ decays,
 since they are Cabibbo suppressed and absent in the $D^{\pm}$ and $D^0$ decays, respectively.
The WA effects compete with the  ones  originated by $SU(3)$ breaking effects in the matrix elements of operators that contribute significantly to the total rates.
At low $q^2$, the WA contributions start at order $1/m_c^3$, while  at high $q^2$ 
 numerically sizable WA contributions to
inclusive rates may be expected  from nonperturbative dynamics,  overcoming the helicity suppression \cite{Bigi:1993bh}.
  An analysis based on inclusive semileptonic $D$ decays, which considers
both the widths and the lepton energy moments, shows no clear evidence of WA effects, i.e. the description in terms of operator product expansion reproduces  well the experimental  data \cite{Gambino:2010jz}.
Based on the previous results,  we neglect possible WA contributions  in   Eq. \ref{rough1}  at $q^2=0$ and at the current level of precision.
While WA might affect the corresponding
inclusive semileptonic width only moderately, it should impact the exclusive channels
$D_s^+ \to \etp l^+\nu $ and  $D^+ \to  \etp l^+\nu $ on
the Cabibbo-favored and suppressed levels via the $\etp$'s gluonic component. The strength of the
effect depends on two factors, namely, the size of the $gg$ component in the $\eta^{\prime}$
wave function and on how much $gg$ radiation one can expect in semileptonic $D_s^+$ and  $D^+$ decays. Last, since the main effect might come from the interference with the spectator amplitude, it
can {\it  a priori}  enhance or reduce those rates. 
The  form factor values  at high $q^2$  on lattice have not yet been determined directly \cite{Bali:2011yx, Koponen:2012fu}, but have been extrapolated using  the  single pole model  and the z-parametrization, respectively.
In the following, we  make the reasonable assumption that   small WA additions only affect the size of the leading amplitude through mixing, in a percentage that is within
 the errors we already consider.
We  maintain the possibility of additional contributions from gluonic components only in the form of the simple parametrization 
\ref{colla1}.

By using the CLEO first absolute
measurement of the branching fractions of $ {\cal{B}} ( D_s^+ \rightarrow \eta^{(\prime )} e^+ \nu)$ \cite{:2009cm}, we can write the  ratio
\beq
\left. \frac{ {\cal{B}}(D_s^+ \rightarrow  \eta^\prime e^+ \nu_e)}{{\cal{B}}(D_s^+ \rightarrow  \eta e^+ \nu_e)}
\right|_{\rm CLEO} = 0.36 \pm 0.14
\label{newCleodata}
\eeq
which corresponds to
\beq
\frac{\Gamma(D_s^+  \to \eta^{\prime} e^+ \nu_e)}{\Gamma(D_s^+  \to \eta \, e^+ \nu_e)} = \frac{\int_0^{(m_{D_s}-m_{\eta^\prime})^2} dq^2 \,  |\bar p_{\eta^\prime}(q^2) |^3 \, { | F_1(q^2)^{D_s \ra \eta_s}|}^2
}{\int_0^{(m_{D_s}-m_\eta)^2}  dq^2 \, |\bar p_\eta (q^2) |^3 \, { | F_1(q^2)^{D_s \ra \eta_s}|}^2}\cot ^2 \phi \cos^2\phi_G
\label{withgluon}
\eeq
To calculate the explicit form of the ratio one has to model the $q^2$ dependence of the form factors; we have considered the lattice points  for $F_1^{D_s \ra \eta_s}$, as plotted in \cite{Koponen:2012fu}, and a model extrapolation to the full range. 
 The errors have been taken conservatively,  as the maximum errors read on the plot, in both the coarse and  fine lattices. 
 The assumption of pole dominance seems to be more tenable in the case of $D$ than of $B$ decays, since the $q^2$ range is smaller, a few GeV against about 25 GeV.
Our fit has been performed  on  the simple pole model, where
where a single pole dominance is assumed. By restricting to  the form factor $F_1(q^2)$,  we have 
\beq
F_1(q^2)= \frac{F_1(0)}{1-\frac{q^2}{m_{\mathrm{pole}}^2}}
\label{singlepole1}
\eeq
We have also performed the fit on
the modified pole model
\cite{Becirevic:1999kt}, where
\beq
F_1(q^2)= \frac{F_1(0)}{ \left(1-\frac{q^2}{m_{\mathrm{pole}}^2}\right)  \left(1- \alpha\frac{q^2}{m_{\mathrm{pole}}^2}\right)}
\label{doublepole1}
\eeq
In the single pole model, our estimate gives $m_{pole}= 1.88 \pm 0.02$. This value is not very different from the value estimated by  CLEO  \cite{Besson:2009uv}  in the $D^+ \ra \bar K^0 e^+ \nu_e$ case, that is $m_{pole}= 1.95 \pm 0.03 \pm 0.01$, where the errors are statistical and systematic. Such similarity was expected, since it is a  consequence of the approximate independence on the spectator quark  recently exposed by the lattice result  $F_1^{D_s \ra \eta_s}(q^2) \sim  F_1^{D^+ \ra K^0} (q^2)$ within 3\% \cite{Koponen:2012fu}. 
In the case of the  modified pole model, we estimate  $\alpha = 0.36 \pm 0.04$, to be compared with $\alpha= 0.28 \pm 0.06 \pm 0.02 $   in the $D^+ \ra \bar K^0 e^+ \nu_e$ decay  \cite{Besson:2009uv}.

Since Eq. \ref{withgluon}  is affected, even in this simple approach, by the gluonic component of the $\eta^\prime$ wave function, we prefer to consider  a slightly different ratio, the one between Cabibbo-favored and Cabibbo-suppressed widths, which  is independent of the angle  $\phi_G$
 \bea
 \frac{\Gamma(D_s^+  \to \eta^{\prime} e^+ \nu_e)/\Gamma(D_s^+  \to \eta \, e^+ \nu_e)}{\Gamma(D^+ \to \eta^{\prime} e^+ \nu_e)/\Gamma(D^+ \to \eta \, e^+ \nu_e)} 
&=&   \frac{\int_0^{(m_{D_s}-m_{\eta^\prime})^2} dq^2  |\bar p_{\eta^\prime}(q^2) |^3  { | F_1(q^2)^{D_s \ra \eta_s}|}^2}{ \int_0^{(m_{D_s}-m_\eta)^2}  dq^2  |\bar p_\eta (q^2) |^3 { | F_1(q^2)^{D_s \ra \eta_s}|}^2}
\times \nonumber \\ 
&\times & \frac{\int_0^{(m_{D}-m_{\eta})^2} dq^2   |\bar p_\eta(q^2) |^3  { | F_1(q^2)^{D \ra \eta_q}|}^2}{ \int_0^{(m_{D}-m_{\eta^\prime})^2}  dq^2  |\bar  p_{\eta^\prime} (q^2) |^3  { | F_1(q^2)^{D\ra \eta_q}|}^2} \cot^4 \phi
\label{doubleratio}
 \eea

Last year,  the  CLEO collaboration presented
the first observation of $ D^+ \rightarrow \eta^\prime e^+ \nu$,
with  branching fraction ${\cal{B}}  (D^+ \to \etp e^+ \nu) = (2.16 \pm 0.53 \pm 0.07) $ x $  10^{-4}$,
and an improved ${\cal{B}} (D^+ \rightarrow \eta e^+ \nu) =
(11.4 \pm 0.9 \pm 0.4)$ x $ 10^{-4}$ \cite{Yelton:2010js}.
From  data, we get
\begin{equation}
\left. \frac{{\cal{B}}(D^+ \to \eta^{\prime} e^+ \nu)}{{\cal{B}}(D^+ \to \eta e^+ \nu)}\right|_{\mathrm{CLEO}} = 1.9 \pm 0.9
\label{ratiowidth2}
\end{equation}
The sizable error is because of the large error in \ref{newCleodata}, which in turn follows from the large statistical uncertainty in what is the first determination of 
${\cal{B}}( D^+ \ra \eta^\prime e^+ \nu_e $). The statistical error is expected to reduce with future data; by a reduction  of just one-half, the error in \ref {ratiowidth2} shrinks to about  20\%.
In \ref{doubleratio},  we consider   the same  $F_1^{D_s \ra \eta_s} (q^2)$   employed in Eq. \ref{withgluon}.
No direct lattice points are available at the moment for $F_1^{D \ra \eta_q}(q^2)$; still  the lattice provides the interesting result of the  independence from the spectator quark
  $F_1^{D_s \ra K}(q^2) \sim  F_1^{D \ra \pi} (q^2)$ \cite{Koponen:2012fu}. All these processes are Cabibbo suppressed 
 and single out the nonstrange components of the wave functions; therefore, it  is reasonable to assume  $ F_1^{D \ra \pi} (q^2) \sim  F_1^{D \ra \eta_q}(q^2)$. We include the small dependence 
on the spectator quark in the theoretical error, and extrapolate to the full $q^2$ range by using the simple and the modified pole models. The parameter estimates give $m_{pole}= 1.9 \pm 0.2$ and $\alpha=0.21 \pm 0.04$. By comparing Eq. \ref{doubleratio} with the experimental data, we obtain
\beq
\phi= (41 \pm 3)^\circ
\label{result2}
\eeq
To be conservative, we have used  the different parametrizations in $q^2$ as an additional maximum theoretical error,
 but its effects, as the effect of the theoretical errors in the parameters of the models, are negligible with respect to the experimental error in \ref{newCleodata}.
New expected data will allow a substantially more accurate determination.
%
%
The agreement with other determinations from semileptonic decays based on different phenomenological approaches and older data   is remarkable \cite{Feld&S, GROROS, Anisovich:1997dz, DiDonato:2011kr}.
The agreement extends also to extractions from other strong and electroweak processes at lower energy, as can be seen in Table \ref{phidectab2}.
\begin{table}[t]
\centering
\vskip 0.1 in
\begin{tabular}{|l|c|} 
\hline
Decays &    $\phi$\\
\hline
 $\phi/\rho/\omega  \to \gamma \eta^{({\prime})} $ + $\eta^{\prime} \to    \gamma \, \rho/\omega  $ \cite{Rafel}  &  $(41.4\pm1.3)^{\circ}$ \\
$\phi/\rho/\omega  \to \gamma \eta^{({\prime})} $ + $\eta^{\prime} \to    \gamma \, \rho/\omega $  + $\pi^0/\eta^{\prime} \to  \gamma \gamma$ \cite{Ambrosino:2009sc}  & $(40.4\pm0.6)^{\circ}$\\
$\psi \to \rho /\omega /\phi \,  \eta^{({\prime})}  $ \cite{RafelJPsi} &  $(44.6\pm 4.4)^{\circ}$ \\
$\psi \to \rho /\omega /\phi \,  \eta^{({\prime})}  $ \cite{Thomas} &  $ (46^{+4}_{-5})^{\circ}$ \\
$D_s^+  \to \eta^{(\prime)} e^+ \nu  $ \cite{Anisovich:1997dz} &  $ (37.7 \pm 2.6)^{\circ} $ \\
$D_{(s)}^+  \to \eta^{(\prime)} e^+ \nu  $ (this work) &  $ (41 \pm 3)^{\circ} $ \\
\hline
\end{tabular}
\caption{ \it Comparison between some determinations of the mixing angle $\phi$, obtained in different processes and approaches,  all of them allowing for a
 nonzero gluonium component.}
\label{phidectab2}
\end{table}
%
By combining the result \ref{result2} with Eq. \ref{withgluon} one can at the most obtain an upper limit on the mixing angle $\phi_G$, which is about  $40^\circ$, but any  conclusion on the gluonic content is prevented by
the large size of the actual  experimental error. The precision of the estimate will benefit from expected new data; more statistics will also allow, within the same approach, a precise estimate of the gluonic mixing angle.

Final further remarks are in order. In the vector sector, we do not expect the  mixing  $\phi$--$\omega$ to be as large as in the pseudoscalar sector,   because there is no additional mixing induced  by the axial $U(1)$ anomaly.
In the absence of mixing,   the state $\omega $ has no strange valence quark and  corresponds to  $|u \bar u + d \bar d\rangle/\sqrt 2 $. 
$D_s$  Cabibbo-favored semileptonic decays are expected to lead to final states that
can couple to $|\bar s s \rangle$, in the quark flavor basis. The  decay $ D^+_s \rightarrow \omega e^+ \nu_e $
 occurs through $\phi$--$\omega$ mixing and/or  WA diagrams, where  the lepton pair couples weakly to the $c \bar s$ vertex. In the hypothesis 
of WA dominance and using factorization, the corresponding branching fraction was estimated to be
$ (0.13 \pm 0.05) \%$ \cite{Gronau:2009mp}.
Experimentally,
only an upper limit is available at the moment,
limiting the  branching fraction
to less than  0.20\% at 90\% C.L.
\cite{Martin:2011rd}.

Semileptonic $D$ decays also offer the chance to explore possible exotic states.
An interesting channel is the  $D^+_s \rightarrow f_0(980) \, l^+ \nu $ decay, where  experimental results  have been provided for the first time  by CLEO in 2009 
\cite{Ecklund:2009aa}.
The 
   nontrivial nature
of the experimentally  well-established $f_0(980)$ state has been discussed
for decades and there are still different interpretations, from the conventional quark-antiquark picture,  to multiquark  or molecular
bound states.  The channels $D_{(s)}^+ \rightarrow f_0(980) \, l^+ \nu $ 
  can be used as a probe on  the hadronic structure  of the light scalar resonance; for recent studies see \cite{ Wang:2009azc, Fariborz:2011xb,  Achasov:2012kk, Pennington:2010dc}.
A further handle is given by the possibility  to correlate observables related to the charm  semileptonic branching ratios with theoretical and experimental analyses of the  hadronic $B_s \ra J/\psi f_0$ decay
\cite{Stone:2008ak, Fleischer:2011au}.

\vspace{0.5cm}
We thank Issaku Kanamori for interesting discussions.



\begin{thebibliography}{99}

\bibitem{DiDonato:2011kr}
  C.~Di Donato, G.~Ricciardi and I.~I.~Bigi,
  Phys.\ Rev.\ D {\bf 85} (2012) 013016.

\bibitem{MyBReport}
  G.~Ricciardi,
  Mod.\ Phys.\ Lett.\ A {\bf 27} (2012) 1230037.









\bibitem{Fleischer:2011ib} 
  R.~Fleischer, R.~Knegjens and G.~Ricciardi,
  Eur.\ Phys.\ J.\ C {\bf 71} (2011) 1798. 

\bibitem{cleo}
G.~Brandenburg et al. [CLEOII Collaboration],
 Phys. Rev. Lett. {\bf 75} (1995) 3804.

\bibitem{:2009cm}
  J.~Yelton et al. [CLEO Collaboration],
  Phys. Rev.  D {\bf 80} (2009) 052007.


\bibitem{Mitchell:2008kb}
  R.~E.~Mitchell {\it et al.}  [CLEO Collaboration],
  Phys.\ Rev.\ Lett.\  {\bf 102} (2009) 081801.

\bibitem{Yelton:2010js} 
  J.~Yelton {\it et al.}  [CLEO Collaboration],
  Phys.\ Rev.\ D {\bf 84} (2011)  032001.


 \bibitem{DeFazio:2000my}
  F.~De Fazio and M.~R.~Pennington,
  JHEP {\bf 0007} (2000) 051.

\bibitem{Christ:2010dd}
  N.~H.~Christ
{\it et al.},
  Phys.\ Rev.\ Lett.\  {\bf 105} (2010) 241601.

\bibitem{Dudek:2011tt}
  J.~J.~Dudek, R.~G.~Edwards, B.~Joo, M.~J.~Peardon, D.~G.~Richards and C.~E.~Thomas,
  Phys.\ Rev.\ D {\bf 83} (2011) 111502.

\bibitem{McNeile:2000hf}
  C.~McNeile {\it et al.}  [UKQCD Collaboration],
  Phys.\ Lett.\ B {\bf 491} (2000) 123
   [Erratum-ibid.\ B {\bf 551} (2003) 391].


\bibitem{Ottnad:2012fv}
  K.~Ottnad {\it et al.}  [ETM Collaboration],
 JHEP {\bf 1211} (2012) 048.



\bibitem{Bali:2011yx}
  G.~S.~Bali {\it et al.}  [QCDSF Collaboration],
  PoS LATTICE {\bf 2011} (2011) 283.

\bibitem{Kanamori}
  I. Kanamori
( private 
communication).


\bibitem{Gregory:2011sg}
  E.~B.~Gregory, A. C. Irving, C. M. Richards and C. McNeile   [UKQCD Collaboration],
  Phys.\ Rev.\ D {\bf 86} (2012) 014504.

\bibitem{Colangelo:2001cv}
  P.~Colangelo and F.~De Fazio,
  Phys.\ Lett.\ B {\bf 520} (2001) 78.

\bibitem{Koponen:2012fu}
  J.~Koponen, C.~T.~H.~Davies and G.~Donald,
 in Proceedings of the 5th International Workshop on Charm Physics (Charm 2012),  arXiv:1208.6242 [hep-lat].

\bibitem{Becirevic:1999kt}
  D.~Becirevic and A.~B.~Kaidalov,
Phys.\ Lett.\ B {\bf 478} (2000) 417.

\bibitem{Besson:2009uv}
  D.~Besson {\it et al.}  [CLEO Collaboration],
  Phys.\ Rev.\ D {\bf 80} (2009) 032005.

\bibitem{Bigi:1993bh}
  I.~I.~Y.~Bigi and N.~G.~Uraltsev,
  Nucl.\ Phys.\ B {\bf 423} (1994) 33.

\bibitem{Asner:2009pu}
  D.~M.~Asner {\it et al.}  [CLEO Collaboration],
  Phys.\ Rev.\ D {\bf 81} (2010) 052007.


\bibitem{Bianco:2003vb}
  S.~Bianco, F.~L.~Fabbri, D.~Benson and I.~Bigi,
  Riv.\ Nuovo Cim.\  {\bf 26N7} (2003) 1.


\bibitem{Gambino:2010jz}
  P.~Gambino and J.~F.~Kamenik,
  Nucl.\ Phys.\ B {\bf 840} (2010) 424.


\bibitem{Feld&S}
T.~Feldmann, P.~Kroll and B.~Stech, 
Phys. Rev. D {\bf 58} (1998) 114006.

\bibitem{GROROS}
M. Gronau, J.L. Rosner, 
 Phys. Rev. D
{\bf 83} (2011) 034025.

\bibitem{Anisovich:1997dz}
  V.~V.~Anisovich, D.~V.~Bugg, D.~I.~Melikhov, V.~A.~Nikonov,
Phys. Lett. B
 {\bf 404} (1997) 166.


\bibitem{Ambrosino:2009sc}
  F.~Ambrosino et al. [KLOE Collaboration],
  JHEP {\bf 0907} (2009) 105.


\bibitem{Rafel}R.~Escribano and J.~Nadal, 
JHEP {\bf 05} (2007) 006.

\bibitem{RafelJPsi} R.~Escribano, 
  Eur.\ Phys.\ J.\  {\bf C65} (2010) 467.

\bibitem{Thomas} C.~E.~Thomas,
  JHEP {\bf 0710} (2007) 026.

\bibitem{Gronau:2009mp}
  M.~Gronau and J.~L.~Rosner,
  Phys.\ Rev.\ D {\bf 79} (2009) 074006.


\bibitem{Martin:2011rd}
  L.~Martin {\it et al.}  [CLEO Collaboration],
  Phys.\ Rev.\ D {\bf 84} (2011) 012005.



\bibitem{Ecklund:2009aa}
  K.~M.~Ecklund {\it et al.}  [CLEO Collaboration],
  Phys.\ Rev.\ D {\bf 80} (2009) 052009.



\bibitem{Wang:2009azc}
  W.~Wang and C.-D.~Lu,
  Phys.\ Rev.\ D {\bf 82} (2010) 034016.

\bibitem{Pennington:2010dc}
  M.~R.~Pennington,
  AIP Conf.\ Proc.\  {\bf 1257} (2010) 27.


\bibitem{Fariborz:2011xb}
  A.~H.~Fariborz, R.~Jora, J.~Schechter and M.~Naeem Shahid,
  Phys.\ Rev.\ D {\bf 84} (2011) 094024.


\bibitem{Achasov:2012kk}
  N.~N.~Achasov and A.~V.~Kiselev,
Phys. Rev. D {\bf 86} (2012) 114010.

\bibitem{Stone:2008ak}
  S.~Stone and L.~Zhang,
  Phys.\ Rev.\ D {\bf 79} (2009) 074024.

\bibitem{Fleischer:2011au}
  R.~Fleischer, R.~Knegjens and G.~Ricciardi,
  Eur.\ Phys.\ J.\ C {\bf 71} (2011) 1832.








\end{thebibliography}
\end{document}